\documentclass[a4paper]{article}
\usepackage[latin1]{inputenc} 
\usepackage[T1]{fontenc} 
\usepackage{RR,RRthemes} 
\usepackage{hyperref}
\usepackage{listings}     
\usepackage{color} 
\usepackage{bera}  

\usepackage{amsmath}
\usepackage{paralist}
\usepackage{url}
\usepackage{amssymb}
\usepackage{xspace}
\usepackage{graphicx}
\usepackage{calc}  
\usepackage{macros} 
\newsavebox\loader
\newsavebox\cmd
\newsavebox\evt
\newsavebox\sync

\lstdefinelanguage[bzr]{c++}
 {basicstyle=\scriptsize,
    morekeywords={node, returns, let, tel, peId, peid,  int,  var, contract,
      assume, enforce, with, bool, *, +, if, then , else, hwParam,
      func, main, and, not, until, state, do, true, false, automaton, end},  
 }

 \lstdefinelanguage{idl2}
 {
  basicstyle=\scriptsize,
  morekeywords={in, out, interface, void},
  backgroundcolor=\color{yellow},
 }

\RRdate{May 2011}

\RRauthor{

 Tayeb Bouhadiba   \and
 Quentin Sabah   \and
Gwenaël Delaval   \and
Eric Rutten   
}
\RRtitle{Synchronous Control
of Reconfiguration in Fractal Component-based Systems
 -- a Case Study}
\RRetitle{Synchronous Control 
of Reconfiguration in Fractal Component-based Systems
 -- a Case Study}
\titlehead{BZR for the control of reconfigurations in Fractal}
\RRresume{ Dans le contexte des composants pour systèmes embarqués, la
  gestion de la reconfiguration dynamique devient de plus en plus
  importante. Le modèle à composants Fractal et son implémentation
  MIND, fournissent des moyens de contrôle de cycle de vie des
  composants ainsi que des moyen pour le contrôle des
  architectures. L'utilisation des architectures intégrées de plus en
  plus complexes, rend la gestion des opérations de reconfiguration
  difficile à maintenir par le programmeur. Cette gestion devient plus
  complexe quand des propriétés globales sur le systèmes doivent être
  assurées. 

  Nous proposons d'utiliser des langages synchrones réactifs, reposant
  sur des modèles comportementaux sous la forme de systèmes de
  transitions. De plus, notre approches, qui produit un manager
  synchrone pour la reconfiguration dynamique profite des techniques
  formelles comme la Synthèse de Contrôleurs Discrets.

  Ce papier décrit l'intégration concrète d'un manager synchrone pour
  la reconfiguration de systèmes-à-composants Fractal. Nous
  détaillerons notre approche en commençant par la partie modélisation
  du problème de contrôle sous forme d'espace d'états de
  configurations, ainsi que la description des propriétés de
  contrôle. Ensuite, nous aborderons la partie implémentation du
  manager résultant en Fractal/Cecilia et son intégration dans des
  applications Fractal distribuées en utilisant le middleware
  Comete. Nous validerons notre approche au moyen d'un cas d'étude sur
  le serveur HTTP Comanche sur une plateforme d'exécution
  multicoeurs. }

\RRabstract{ In the context of component-based
  embedded systems, the management of dynamic reconfiguration in
  adaptive systems is an increasingly important feature.  The Fractal
  component-based framework, and its industrial instantiation MIND,
  provide for support for control operations in the lifecycle of
  components.  Nevertheless, the use of complex and integrated
  architectures make the management of this reconfiguration operations
  difficult to handle by programmers. To address this issue, we
  propose to use synchronous languages, which are a complete approach
  to the design of reactive systems, based on behavior models in the
  form of transition systems.  Furthermore, the design of closed-loop
  reactive managers of reconfigurations can benefit from formal tools
  like Discrete Controller Synthesis.
  In this paper we describe an approach to concretely integrate
  synchronous reconfiguration managers in Fractal component-based
  systems.  We describe how to model the state space of the control
  problem, and how to specify the control objectives.  We describe the
  implementation of the resulting manager with the Fractal/Cecilia
  programming environment, taking advantage of the Comete distributed
  middleware.  We illustrate and validate it with the case study of
  the Comanche HTTP server on a multi-core execution platform.
}
\RRmotcle{Systèmes à base de composants, Programmation Synchrone,
  Systèmes reconfigurables, Synthèse de contrôleurs discrets.}
\RRkeyword{Component-based systems,
synchronous programming,
reconfigurable systems,
discrete controller synthesis.}
 \RRprojet{SARDES}  
\RRdomaine{Distributed Systems and Services} 
\RRthemeProj{Sardes} 
\RRdomaineProjBis{Sardes} 
\RCGrenoble 

\begin{document}
\RRNo{7631} 
\makeRR   


\tableofcontents
\newpage
\newcommand{\todo}[1]{{\tt {\color{red} TODO or Replace}:  #1}}
\newcommand{\redo}[1]{{\tt {\color{red}#1}}}

\lstdefinelanguage[bzr]{c++}
 {basicstyle=\scriptsize,
    morekeywords={node, returns, let, tel, peId, peid,  int,  var, contract,
      assume, enforce, with, bool, *, +, if, then , else, hwParam,
      func, main, and, not, until, state, do, true, false, automaton, end},  
 }

 \lstdefinelanguage{idl}
 {
  basicstyle=\scriptsize,
  morekeywords={in, out, interface}
 }

\newcommand{\fractal}{{\sc Fractal}\xspace}
\newcommand{\cecilia}{{\sc Cecilia}\xspace}
\newcommand{\Comete}{{\sc Comete}\xspace}
\newcommand{\MIND}{{\sc MIND}\xspace}


\newcommand{\filehandlerone}{{\tt fileserver$_1$}\xspace} 
\newcommand{\filehandlertwo}{{\tt fileserver$_2$}\xspace} 
\newcommand{\logger}{{\tt logger}\xspace} 
\newcommand{\frontend}{{\tt frontend}\xspace} 
\newcommand{\dispatcher}{{\tt dispatcher}\xspace} 
\newcommand{\analyzer}{{\tt analyzer}\xspace} 

\newcommand{\Srun}{{\tt R}\xspace} 
\newcommand{\Sunbound}{{\tt SS}\xspace} 
\newcommand{\Sstop}{{\tt S}\xspace} 
\newcommand{\Sempty}{{\tt FE}\xspace} 
\newcommand{\Sfull}{{\tt FF}\xspace} 
\newcommand{\true}{{\tt 1}\xspace} 
\newcommand{\false}{{\tt 0}\xspace} 
\newcommand{\Sa}{{\tt S$_0$}\xspace} 
\newcommand{\Sb}{{\tt S$_1$}\xspace} 

\newcommand{\Sdone}{{\tt D}\xspace} 
\newcommand{\Spend}{{\tt P}\xspace} 

\newcommand{\Ssd}{{\tt \sc s,d}\xspace} 
\newcommand{\Ssp}{{\tt \sc s,p}\xspace} 
\newcommand{\Srp}{{\tt \sc r,p}\xspace} 
\newcommand{\Srd}{{\tt \sc r,d}\xspace} 
\newcommand{\Sssd}{{\tt \sc ss,d}\xspace} 
\newcommand{\Sssp}{{\tt \sc ss,p}\xspace}

\newcommand{\evtreceiver}{{\tt evt}\xspace} 
\newcommand{\synctrl}{{\tt ctrl}\xspace} 
\newcommand{\cmdmanager}{{\tt cmd}\xspace} 

\newcommand{\syncprog}{manager\xspace}
\newcommand{\syncctrl}{controller\xspace} 

\section{Introduction}


In the context of component-based embedded systems, the management of
reconfiguration in adaptive systems is an increasingly important
feature. The Fractal component-based framework,
  and its industrial instantiation MIND, provide for support
for control operations in the lifecycle of components.
Nevertheless, the use of complex and integrated architectures 
make the management of this reconfiguration operations difficult to handle by
programmers. To address this issue, we propose to use
synchronous languages, which are a complete approach to the design of
reactive systems, based on behavior models in the form of transition
systems.  Furthermore, the design of closed-loop reactive controllers
of reconfigurations can benefit from formal tools like Discrete Controller Synthesis (DCS).

Using DCS, integrated in a programming language, provides designers for support
in the correct design of controllers. This method is different from the usual
method of first programming and then verifying. It involves the automated
generation of part of the control logic of a system. Discrete control has until
now been applied to computing systems only very rarely \cite{wang-wodes10}.
An important challenge is to 
integrate this formal reactive systems design in actual, practical operating systems. 
An open issue is the identification and correct use of 
the practical sensors and monitors providing for reliable and significant information;
the control points and actuators available in the API of the OS, enabling enforcement of a management policy;
the firing conditions for the transitions of the automata.

 
In this paper we describe an approach to concretely integrate
synchronous reconfiguration controllers in Fractal component-based
systems.
Our contribution is: (i) a synchronous model of the behavior of the
reconfigurable components, in the form of the state space of the
control problem, and the specification of the control objectives
(Section \ref{sec-ctrl}); (ii) a component-based architecture for the
implementation of the resulting controller with the Fractal/Cecilia
programming environment, taking advantage of the Comete middleware.
(Sections \ref{sec-impl} and \ref{sec-impl-async}). We validate it by
the case study of the Comanche HTTP server deployed on a multi-core
execution platform.

\begin{figure*}
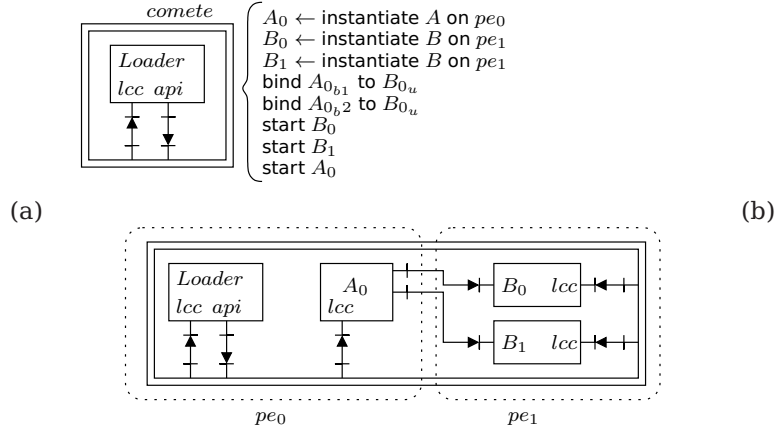

\centering
(a) \scalebox{0.95}{ \input{deploy-one} }
(b) \scalebox{0.95}{  \input{deploy-two} }
\caption{The deployment process.  
(a) the {\em loader} is in charge of building the initial 
configuration 
(b)} 

 \label{fig:deploy}
  \label{fig:deploy-two}
 \label{fig:deploy-one} 
\end{figure*}

\section{Background}

\subsection{The Fractal Component Model}
\label{ss:fractal}

We introduce \fractal \cite{ow2-fractal}\cite{Bruneton02recursiveand},
a hierarchical and reflective component model and \cecilia
\cite{ow2-cecilia}, a component-base software engineering framework
providing a C implementation of this model. \fractal defines
components as entities encompassing behaviours and data.  A component
can be dismentled in two parts: a {\em membrane} and a {\em content}.
The content is either a set of operations or a finite number of
sub-components, which are under the control of the enclosing
membrane. Components can be nested at an arbitrary level in a
recursive fashion.

%
%
%

\begin{figure}[b]
\centering
\scalebox{0.6}{\input{fractal-vocabulary-simple}}

\caption{\fractal vocabulary.}
\label{fig:fractal-vocabulary}
\end{figure}

Components interact with their environment through {\em interfaces}. Interfaces
are typed collections of operations. They can be of two sorts: 
{\em client interfaces} emit operation invocations, {\em server interfaces} 
receive operation invocations. One-way operation invocations may carry 
arguments, two-way operations consist of an invocation followed by the return of
a result. 
Components can expose several interfaces. 
Client and server
interfaces are connected through explicit {\em bindings}.
{\em Functional
interfaces} are access points to content operations, while {\em Controller 
interfaces} define membrane operations. The membrane embodies the control
behaviour associated with a particular component.

The \fractal model defines a set of optional controller interfaces to adress
minimal requirements in terms of introspection, composition and life-cycle. 
Among others are:

\begin{compactitem}
\item {\em Life-Cycle Controller}: controls component's behavioural phases
such as starting and stopping.

\item {\em Binding Controller}: establish/break bindings between component's
interfaces and its environment.
\end{compactitem}

The \cecilia framework is a coherent toolchain to design, compile and deploy 
component-based applications. It allows the description of hierarchical
\fractal architectures using an {\sc xml} ADL, and the implementation of the 
primitive components using the {\sc C} langage. By automaticaly generating the 
controllers related glue-code, the toolchain allow the developer to focus on 
the implementation of the primitives' content operations i.e. the {\em functional
interfaces}' methods in figure \ref{fig:fractal-vocabulary}.
From an application definition and its implementation, the toolchain generate
a standalone executable or a set of independent component binaries. The \Comete
middleware described in the next section relies on these independent bricks to 
dynamically load and bind instances of components to build arbitrary 
architectures.

\subsection{Comete}
\label{ss:comete}
\newcommand{\pe}[1]{{\tt pe$_#1$}\xspace} 

\Comete~\cite{ow2-comete} is a
minimal middleware and run-time layer engineered by STMicroelectronics
to dynamicaly deploy, run and reconfigure Cecilia components over
distributed platforms.
\Comete is providing a {\em distributed event-driven architecture}.
Applications deployed using \Comete can take advantage of the
high-level abstraction of this platform to communicate {\em
  asynchronous messages} between components.
The middleware models a distributed platform as a set of {\em
  processing elements} and handles communication between them so that
application developpers don't have to know about the underlying
communication channels and protocols.  


The first step of any application deployment using \Comete lies in the
instantiation of a {\em loader} component (Figure
\ref{fig:deploy-one}(a)) bound to the \Comete API interface (this
interface is further detailed in Listing~\ref{cod:itf-events-comete}).
This mandatory component is in charge of deploying the application by
invoking \Comete operations such as instantiations and bindings to
build the initial configuration (Figure \ref{fig:deploy-one}(b)).
%
%
Once the deployment process done, the loader can remotely manipulate
{\em LifeCycleController (lcc)} interface of any instantiated component through
the {\em start/stop} methods to initiate the execution. Then, the 
application is free to diverge from its initial
architecture by successive reconfigurations.
The initial deployment is like a reconfiguration 
from a single primitive component to a potentially more complex architecture.

\begin{figure}[h]
\centering
\scalebox{0.7}{\input{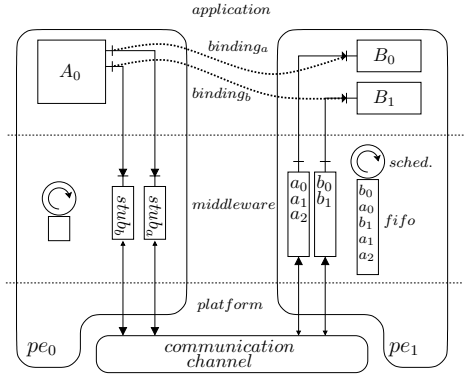}}

\caption{A simple application spanning over two processing elements. Internal 
implementation of applicative bindings (stroked) involves dedicated
communication components in the middleware layer to support platform
channels.}
\label{fig:comete-internals}
\end{figure}

Each processing element executes message handlers related to
components it embodies. The runtime layer consists of a task queue
associated with a FIFO scheduler (Figure
\ref{fig:comete-internals}). The scheduler is non preemptive. Every
message directed to a given component will be handled as task on the
respective processing element. The execution model guaranties that: a)
At any given time only one method is executing on a processing
element. b) Components deployed on different processing element may
execute methods concurrently.


At its lowest level, the runtime layer is dealing with platform heterogeneity
in terms of operating systems, hardware platforms and communication protocols.
This heterogeneity is then abstracted by the middleware layer.
Internally, a remote binding is handled by a couple of {\em stub/skeleton} 
components loaded respectively on the client's and server's processing element. 
On the client side, the stub transparently intercepts and serializes emitted 
messages. Data are then transmitted through a platform channel to the server 
side. The skeleton reads serialized message data, operates the inverse process 
and pushes a handler task in the local runtime queue. 
The message arguments are stored until the associated task is scheduled. 
Thanks to a generic, scalable, component-based architecture, \Comete is 
targeting a wide range of platforms, from embedded System-On-Chip to distributed
computers. The middleware is built over an extensible library of components 
providing support for various processing elements and communication channels.
Noticeable specializations of \Comete are: 
the STm8010(Traviata) board including three ST200 cores, 
the xStream many-cores streaming prototype, 
posix-compliant operating systems easing deployment over multi-threaded hardware, 
computers distributed over a TCP/IP network.

%


\subsection{Heptagon and BZR}



\subsubsection{Heptagon language}
\label{sec:hept-lang}

In this work, we use 
the language
Heptagon/BZR~\cite{delaval10:_contracts_mod_dcs}%
\footnote{Available at
\url{bzr.inria.fr}}.
 The Heptagon language allows to describe
reactive systems by means of generalized Moore machines, i.e., mixed synchronous
dataflow equations and automata~\cite{lucy:emsoft05b}, with parallel and
hierarchical composition.
The basic behavior is that at each reaction step, values in the input flows are
used in order to compute the values in the output flows for that step.  Inside
the nodes, this is expressed as a set of declarations, which takes the form of
equations defining, for each output, the values that the flow takes,
in terms of an expression on other flows' instantaneous values, possibly using values
computed in preceding steps (also known as state values).

Figure~\ref{fig:del-task} shows a small program in this language. It describes the control of a
task, which can either be idle or active. When it is idle, i.e., in the initial
Idle state, then the occurrence of the input \texttt{r} \emph{requests} the
launch of the task. Another input \texttt{c} (which will be controlled further
by the synthesized controller) can either allow the activation, or temporary
block the request and make the automaton go to a waiting state. When active, the
task can be ended with the input \texttt{e}. This
\texttt{delayable} node has two outputs, \texttt{a} featuring the instantaneous
activity of the task, and \texttt{s} being emitted on the instant when it
becomes active:
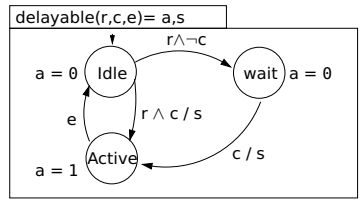
\[
{
  \begin{streams}{8}
    \text{step \#} & 1 & 2 & 3 & 4 & 5 & 6 & 7 & \ldots \\\hline 
    r             & 0 & 1 & 0 & 0 & 0 & 0 & 0 & \ldots \\
    c             & 0 & 0 & 1 & 0 & 0 & 0 & 0 & \ldots \\
    e             & 0 & 0 & 0 & 0 & 0 & 1 & 0 & \ldots \\\hline
    a             & 0 & 0 & 0 & 1 & 1 & 1 & 0 & \ldots \\
    s             & 0 & 0 & 1 & 0 & 0 & 0 & 0 & \ldots \\
  \end{streams}
}
\]

\begin{figure}[h]
 \centering
\scalebox{0.75}{\input{delayable}}
\begin{lstlisting}[language={[bzr]c++}]
node delayable(r,c,e: bool) returns (a,s: bool)
  let
    automaton 
      state Idle
        do a = false ; s = r and c 
        until r and c then Active
            | r and not c then Wait
      state Wait
        do a = false ; s = c 
        until c then Active
      state Active
        do a = true ; s = false
        until e then Idle
     end
   tel
\end{lstlisting}

 \caption{Delayable task (graphical/textual syntax).}
 \label{fig:del-task}
\end{figure}

\subsubsection{BZR and controller synthesis}
\label{sec:bzr-contr-synth}

BZR is an extension of Heptagon, allowing its compilation to involve
\emph{discrete controller synthesis} (DCS)
using the DCS tool \textsc{Sigali}~\cite{marchand00c}.
 DCS allows to compute automatically
a controller, i.e., a function which will act on the initial program so as to
enforce a given temporal property. Concretely, the BZR language allows the
declaration of \emph{controllable variables}, which are not
defined by the programmer. These free variables can be used in the program so as
to let some choices undecided (e.g., choice between several transitions).
The controller, computed by DCS, is then able to avoid undesired
states of the application by setting and updating appropriate values
for these variables at runtime.

Figure~\ref{fig:mut-excl} shows an example of use of these controllable
variables. It consists in two instances of the \texttt{delayable}
node, as in Figure~\ref{fig:del-task}. They run in parallel,
defined by synchronous composition: one global step corresponds to one local
step for every equation, i.e., here, for every instance of the automaton in the
\texttt{delayable} node. Then, the \texttt{twotasks} node so defined is given a
\emph{contract} composed of two parts: the \With part allowing the declaration
of controllable variables ($c_1$ and $c_2$), and the \Enforce part allowing the
programmer to assert the property to be enforced by DCS, using the controllable
variables. Here, we want to ensure that the two tasks running in parallel won't
be both active at the same time. Thus, $c_1$ and $c_2$ will be used by the
computed controller to block some requests, leading automata of tasks to the
waiting state whenever the other task is active.

\begin{figure}[t]
  \centering
  \small
\[
\begin{array}{|l|c}
  \cline{1-1}
  \mathtt{twotasks}(r_1,e_1,r_2,e_2) = a_1,s_1,a_2,s_2 & \\
  \cline{1-1}
  \Enforce \Not (a_1 \And a_2) & \\
  \hline
  \With c_1,c_2 & \multicolumn{1}{c|}{} \\
  \cline{1-1}
  \multicolumn{2}{|c|}{
    \begin{array}{l}
      ~\\
      (a_1,s_1) = \mathtt{delayable}(r_1,c_1,e_1)\\[1ex]
      (a_2,s_2) = \mathtt{delayable}(r_2,c_2,e_2)\\[1ex]
    \end{array}
  }\\
  \hline
\end{array}
\]
   \caption{Mutual exclusion 
 in BZR.}
  \label{fig:mut-excl}
\end{figure}

\subsubsection{Heptagon/BZR compilation}
\label{sec:hept-comp}

The compilation of a Heptagon/BZR program produces sequential code in
a target general programming language (C, Java or Caml). This code
takes the form of two functions (or methods), named \emph{reset} and
\emph{step}. \emph{reset} initializes the internal state of the
program. The \emph{step} function is evaluated at each logical instant
to compute output values from an input vector and the current state,
possibly updated. A typical way of using these functions is to enclose
this \emph{step} call in an infinite loop: 
\begin{alltt}
current_state \(\leftarrow\) \textbf{reset}() \emph{// state initialization}
for each step do
   inputs \(\leftarrow\) gather current events
   outputs, next_state \(\leftarrow\) \textbf{step}(inputs, current_state)
   handle outputs
   current_state \(\leftarrow\) next_state
\end{alltt}

Eventually, such infinite loop will not be as clearly stated, but hidden within,
e.g., events managers, threads, interrupts, depending on the application
context.
In our context, we will use the C generated code. The API, for a node \texttt{f}
with inputs $\ton{\mathtt{x}}{,}$ (typed $\mathtt{t}_i$) and outputs
$\ton[1][p]{\mathtt{y}}{,}$ (typed $\mathtt{t}'_i$) is given below. The
additional input \texttt{mem} is a pointer towards the internal state, to be
used and updated. The result of the \texttt{f\_step} call is a structure in
which are placed the outputs values.


\begin{alltt}
\textkw{void} f_reset(\textkw{f_mem*} mem);
\textkw{f_res} f_step(\textkw{\(\mathtt{t}_1\)} \(\mathtt{x}_1\), ...,\textkw{\(\mathtt{t}_n\)} \(\mathtt{x}_n\), \textkw{f_mem*} mem);
\end{alltt}

\subsection{System/manager Interaction}
\label{ss:interaction}
Our approach applies to systems supporting dynamic reconfiguration and
providing some events describing their observable states. It relies on
the modeling of the system (and possibly the environment) by means of
Heptagon automata and describing the control objectives. The BZR
compilation tool-chain compiles the automata and the control
objectives into a synchronous program which will be referred to as
\syncprog in the sequel. The \syncprog encapsulates a model of the
system and a \syncctrl to enforce the objectives.

\begin{figure}[h]
\centering
\input{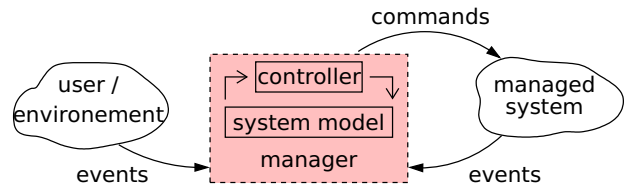}
\caption{Application control by a manager}
\label{fig:control-loop}
\end{figure}

Figure~\ref{fig:control-loop} illustrates the use of a manager (i.e.,
synchronous program) to manage the reconfiguration of a system. The
\syncprog receives some input events from the system and the
environment. According to these inputs and the current state of the
model, the \syncctrl may update the state of the model. Internally,
this reduces to fill the controllable variables with the appropriate
computed values. The computation of the new state of the model
corresponds to a step of the \syncprog and may fire some commands. The
commands are meant to reconfigure the managed system in order for this
latter to be coherent with its model.

\section{The Comanche Http Server}
\label{s:comanche}

This section describes an example of a reconfigurable component-based
software application written in Fractal together with the execution
platform on which the application will be run. The complete system
will be used as a case-study in order to introduce our approach for
the management of reconfigurable systems in Section~\ref{sec-ctrl}.

\subsection{Components architecture}
\begin{figure}[h]
\centering
\input{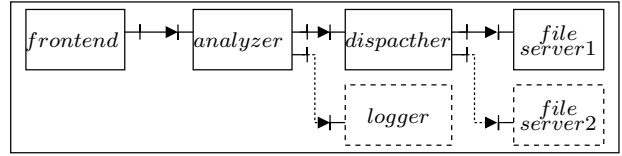}
\caption{Comanche architecture.}
\label{fig:comanche-arch}
\end{figure}

Figure~\ref{fig:comanche-arch} describes the architecture of a HTTP
server written in Fractal. It is a variant of the Comanche server used
as an example in
tutorials\footnote{http://fractal.ow2.org/tutorial}. Incoming requests
are received by the \frontend component, which transmits them to the
\analyzer component. The latter forward well-formed requests to the
\dispatcher which queries \filehandlerone or \filehandlertwo to solve
the requests. The \analyzer component can also send requests to the
\logger 
 to keep track of HTTP requests.

The required components for this application to work are \frontend,
\analyzer, \dispatcher and \filehandlerone. 
A first available degree of
dynamical reconfiguration lies in \filehandlertwo and \logger. As
illustrated by the dashed lines, \filehandlertwo and \logger may be
activated (resp., deactivated) and connected to (resp., disconnected
from) the rest of the components.

\subsection{ Execution architecture}
A second degree of reconfiguration concerns component mapping over
processing elements. For our experiments, we used a server equipped
with two Intel Xeon dual-core processors, each one running at 1.86GHz
clock frequency. The four available cores enable the execution of four
tasks in parallel. In the sequel, we will refer to each core as a {\em
  processing element} (\pe{0}, ..., \pe{3}).

\begin{figure}[h]
\centering
\input{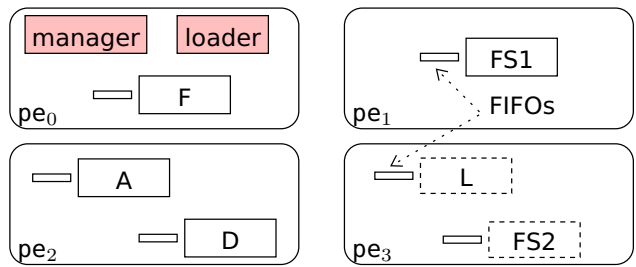}

\caption{Architecture of the controlled application}
\label{fig:global-control-cpt}
\end{figure}

Figure~\ref{fig:global-control-cpt} describes the initial deployment
of Comanche components using Comete middleware
(Section~\ref{ss:comete}) on the execution platform. Each component is
associated with a processing element, on which it will execute in
order to handle messages in its associated FIFO. Component bindings
are implemented by the Comete middleware as asynchronous
communication channels. For the sake of clarity of the figure, the bindings are
made implicit. Notice that the \syncprog and Comete loader also
execute on the same platform.

\subsection{ Renconfiguration policy}
\label{sec-policy}

The control objectives we want to enforce are related to distinct
aspects of the Comanche application and the execution platform. We are
able to design a manager to fire reconfiguration commands in order
to enforce the objectives. Comete together with Fractal API provide
means to reflect manager commands on the managed system. Those means
we are interested in are {\em component migration} provided by Comete
API and {\em lifecycle} (resp., {\em binding}) control interface of
the Fractal API in order to start/stop (resp., bind/unbind)
components. In the following, we list the control objectives:

\begin{description}
\item[Processing element availability] A processing element is a
  shared resource that may be unavailable for some reasons (e.g.,
  energy saving, higher priority task, fault, etc.). We want 
  that no component is running on an unavailable processing element
  ({\em component migration}). 

\item [Workload balancing] There is a maximum workload, a processing
  element should not exceed. Migrate components in order to decrease
  the workload.

\item[Quality-of-Service] When the \filehandlerone is overloaded,
  start the \filehandlertwo in order to keep the average response
  time low ({\em lifecycle and binding control}).

\item[Exclusiveness] The \logger should be started upon the request of
  the user, and then 
 \filehandlertwo should
  not be running ({\em lifecycle and binding control}).
\end{description}

In order to ensure the objectives we described above, a manager is
implemented and put in a closed-loop with the software application as
described in section~\ref{ss:interaction}.
A simple scenario is as follows. The manager receives an
event stating that the \filehandlerone is overloaded. It performs some
computation and emits the command to start the \filehandlertwo and
bind it to the rest of the components. We measure the workload of
\filehandlerone by comparing the size of the FIFO associated to it
with a specific threshold. Next, the controller receives from the
environment that the processing element \pe{3} is no longer
available. The controller reacts by emitting the command to migrate
the components running on \pe{3} to other processing elements, 
balancing the workload as described by the related property.

\section{Designing the manager}
\label{sec-ctrl}

We follow a modeling method for the design of a manager to enforce
the properties we described above; we will elaborate on
it by improving the treament of reconfiguration actions in
Section~\ref{sec-impl-async}.  
The use of DCS ensures that the manager
is correct with respect to the properties given as objectives.
%
It mainly
consists of the following steps: 
(1) Providing a synchronous model of
the behavior of the reconfigurable components and the processing
elements, giving the state space of the control problem;
(2)
Specifying the control objectives and identifying the controllable
variables; (3) 
Compiling: controller synthesis and code generation.

\subsection{Modeling Components With Heptagon}
We adopt a modular modeling approach to enable reusing
models. Moreover, we distinguish between the application models (i.e.,
software) and the execution platform (e.g., hardware)
models. Instances of these models will be associated in the main
synchronous program (see Section~\ref{ss-controller-main}) to
obtain the global model. Hence, the approach facilitates replacing
hardware models without modifying software ones in case the
application is ported to other execution platforms.

\subsubsection{Modeling  Software Components}

\paragraph*{Component Lifecycle}

\begin{figure}[h]
\centering
\scalebox{0.9}{\input{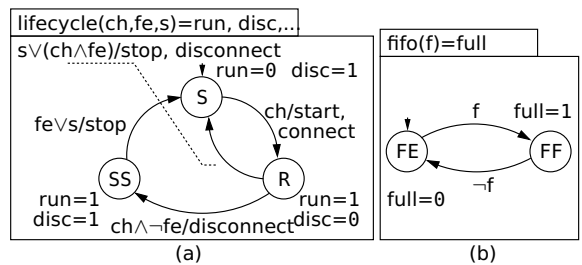}}

\caption{Model of: (a) Lifcecyle, (b)FIFO state}
\label{fig:life-cycle-autom}
\end{figure}

Figure~\ref{fig:life-cycle-autom}-(a) shows the automaton modeling a
Fractal component lifecycle. A component may be in one of three
possible states. A {\tt Running (\Srun)} component is connected to the
other components and may handle incoming messages in its associated
FIFO.  A {\tt Stopped (\Sstop)} component is disconnected from the
others and can not execute to handle messages in its input
FIFO. Finally, a component may be put in a {\tt Safe Stopping
  (\Sunbound)} in order to handle the remaining messages in its FIFO
before completely stopping.

The automaton has three inputs ({\sf ch}, {\sf fe} and {\sf s} for
{\sf change}, {\sf FIFOs empty} and {\sf stop} respectively), and two
outputs {\sf run} and {\sf disc}. Initially, the component is stopped
(i.e., state \Sstop). It goes to \Srun (i.e., running) upon receiving
{\sf ch}. From this state, it goes to state \Sunbound upon receiving
{\sf ch}. At state \Sunbound, it goes to state \Sstop upon receiving
{\sf fe} which means that the input FIFOs are empty. At any state,
receiving {\sf s} forces the automaton to go to the state \Sstop. The
two outputs of the automaton {\sf run} and {\sf disc} (for running and
disconnect) take their values as described by the figure. These
outputs tell whether a component is running (resp., disconnected) or
not.

As described by the transitions of the automaton, some reconfiguration
commands ({\sf start, stop, connect, disconnect}),
represented as additional outputs, are fired upon
changing state. These commands are meant to reconfigure the related
component in order to be coherent with its lifecycle model. For
instance the transition from \Sstop to \Srun, outputs the commands
{\sf start} and {\sf connect} in order to start the component and
connect it to the rest of the components.

\paragraph*{FIFO State}

With each server interface of a component is associated a FIFO. The
FIFO stores the input events before they are handled by the
component. Figure~\ref{fig:life-cycle-autom}-(b) models the state of
such FIFOs. It is a two state automaton with one input and one output
({\sf f} and {\sf full} respectively). Initially, the FIFO is
empty. It goes to the state \Sfull~ (resp., \Sempty) upon receiving
{\sf f} (resp., {\sf $\neg$f}). The value output {\sf full} indicates
the state of the FIFO: true (resp., false) means that the FIFO size is
above (resp., below) a given threshold.

\subsubsection{Modeling Hardware Components}

\paragraph*{Component mapping on Processing Elements}
\newcommand{\Pun}{{\tt PE$_0$}\xspace} 
\newcommand{\Pde}{{\tt PE$_1$}\xspace}
\newcommand{\Pt}{{\tt PE$_2$}\xspace} 
\newcommand{\Pq}{{\tt PE$_3$}\xspace} 

\begin{figure}[h]
\centering
\scalebox{0.85}{\input{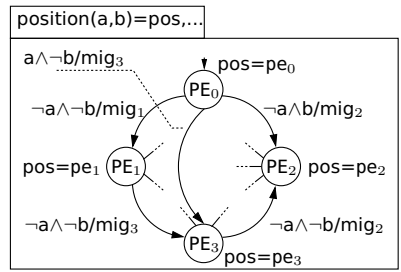} }

\caption{Partial model of a component mapping.}
\label{fig:task-pe-autom}
\end{figure}

Figure~\ref{fig:task-pe-autom} models the mapping of a component on
the available processing elements. It is a four state automaton. Each
state represents mapping on one processing element, on which  a component may run. 
It may change  from one state to another
depending on the Boolean inputs {\sf a} and {\sf b}. That is, each
state has three outgoing transitions (one to each remaining
state). The transition to take depends on the value of {\sf a} and
{\sf b} (two Boolean inputs to encode four possibilities).  For the
sake of clarity, not all of the transitions are present in the
figure. The output of the automaton is of enumerated type; it takes
its values in \{{\pe{0}, \pe{1}, \pe{2}, \pe{3}}\} depending on the
state of the automaton.

The transitions of the automaton are associated with outputs ({\sf mig$_0$, mig$_1$, mig$_2$, mig$_3$}). 
They are associated to 
migration commands, in order to migrate the
corresponding component to its new processing element.

\paragraph*{Processing Element Availability}
\newcommand{\Sdisp}{{\tt ON}\xspace}
\newcommand{\Sndisp}{{\tt OFF}\xspace}

\begin{figure}[h]
\centering
\scalebox{0.85}{\input{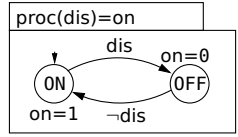}}

\caption{Model of processing element  availability}
\label{fig:core-disp-autom}
\end{figure}

We need a model of the availability of a processing element in order
to know whether a component may run on it or
not. Figure~\ref{fig:core-disp-autom} describes such a model. It is a
two states automaton with one input {\sf dis} (for disable) and one
output {\sf on} which tells on the availability of the processing
element. Initially, a processing element is available (i.e., at state
\Sdisp).  It goes to the state \Sndisp (resp., \Sdisp) upon receiving
{\sf dis} (resp., {\sf $\neg$dis}). The output {\sf on} takes its
value depending on the state of the automaton.

\paragraph*{Processing Element Workload}

The workload of a \pe{i} depends on the components running on it and
their bindings to the other components. Two components $C$ and $S$
(client and server), bound through their respective interfaces $I_c$
and $I_s$ induce some workload on their associated processing elements
\pe{c} and \pe{s}. Based on some benchmarks, we estimate the workload
related to \pe{c} and \pe{s} as described by the following equations:

a) $load(pe,s)=  f \times c$.  \hfill  b) $load(pe,c)=  f \times distance$.

The workload of the processing element running the server component
$S$ depends on the cost $c$ of the executed function, and the function
call frequency $f$. The processing element running the client
component is charged with the communication effort. That is, the
processing element workload depends on the call frequency and a
parameter $distance$ which depends on the position of the two
communicating processing elements. In our case-study, the processing
elements are  processor cores. The communication uses
cache memories of distinct levels. The values we associate to
$distance$ represent the fact that the communication cost differs
depending on the type of cache used between the processing
elements. These values are: $distance=200$ in case \pe{c} and \pe{s}
refer to the same core, $distance=500$ in case the processing elements
are distinct cores of the same processor, $distance=1000$ in case the
two processing elements refer to distinct cores of distinct
processors.

In Listing~\ref{cod:node-cost}, we give the interface of
the node {\rm cost} implementing workload equations. For a given
binding between two components, the node takes as input the position
of the client and the server component, the frequency $f$ of the calls
between them and the cost of the function executed by the server
component. The outputs are integer values associating workloads
induced by the two components with the available processing
elements. The involved processing elements will be associated with
positive integer values, the others will take the value zero.

\begin{lstlisting}[caption={The interface of the node cost}, captionpos=b, label={cod:node-cost},language={[bzr]c++}]
node cost(posC, posS: peid; f, c: int) 
returns (cp0, cp1, cp2, cp3: int)
\end{lstlisting}

\subsection{Complete system model}
\label{ss-controller-main}


The main program (partially described in listing of
Listing~\ref{cod:main-prog}) consists of the synchronous composition
of the automata modeling each component (hardware and software). The
inputs of the main node consist of the events stating on: (pes1) the
availability of the processing element \pe{1}; (add\_L) the request of
the \logger; (f\_S1, f\_S2, f\_L) the load of the FIFO associated with
\filehandlerone, \filehandlertwo and \logger respectively.

The outputs of the main program correspond to the
reconfiguration/migration commands fired by the automata. For instance
mig\_L\_p0 is fired by the automaton modeling the position of the
component \logger (line 22). When mig\_L\_p0 takes the value true, the
component \logger should be migrated to \pe{0}. The output start\_S2
is fired by the automaton modeling \filehandlertwo lifecycle (line
17). When it is true, it states that the component should be started.

The main program is composed of: An instance of node fifo for each
server interface of the components (lines 11-13).  An instance of node
lifecycle for each component (lines 15-17).  An instance of processing
element availability for each processing element (lines 19-20).  An
instance of node position for each component (lines 22-24).  An
instance of node cost for each binding between components (lines
26-28).  Finally, the cost imposed on each processing element
corresponds to the sum of all the cost induced by the bindings (lines
29-30).

\newpage

\lstset{numbers=left,numberstyle=\tiny,stepnumber=2,numbersep=5pt}
\begin{lstlisting}[caption={The main program of the  model}, captionpos=b, label={cod:main-prog},language={[bzr]c++}]
node main(pes1, add_L, f_S1, f_S2, f_L: bool) 
     returns (mig_L_p0, ..., mig_L_p3:int;...               
        mig_F_p0, ..., mig_F_p3:int;
        start_S2, stop_S2, conn_S2, disc_S2:bool;
        start_L, stop_L, conn_L, disc_L:bool)
contract
  enforce(pe_av and wl_ba and qos and exc);
  with(....)

let
  full_S1 = fifo(f_S1); 
  full_S2 = fifo(f_S2);
  full_L  = fifo(f_L); 
  ...
  (run_L,disc_L,start_L,stop_L,conn_L,disc_L)
          = lifecycle(ch_L, full_L, s_L);
  (run_S2,disc_S2...)=lifecycle(ch_S2,full_S2,s_S2);
  ...
  pe1 = proc(pes1);   
  pe2 = proc(false);
  ...
  (pos_L,mig_L_p0,mig_L_p1,mig_L_p2,mig_L_p3)
         = position(a_L, b_L);
  (pos_S1,...)= position(a_S1, b_S1);
  ...
  (c11,c21,c31,c41)=cost(pos_D, pos_S1, f_1, c_1);
  (c12,c22,c32,c42)=cost(pos_D, pos_S1, f_2, c_2);
  ...
  wl1 = c11 + c12 + ... + c18;
  ...
tel 
\end{lstlisting}

\subsection{Describing Control Objectives with BZR}
\label{ss-controller-main}

We now have a complete model of the possible behaviors of the system,
in the absence of control.  We want to obtain a controller that will
enforce the policy given informally in Section~\ref{sec-policy}.  We
do this in the form of a contract for each of the points of the policy
as follows.


\paragraph*{Processing element availability} We want no component running on an unavailable processing element: 
this can be achieved by {\em component migration}.

The corresponding expression in the contract to be enforced i.e., to be controlled for invariance,  is:
\[
{\tt pe\_av =} \bigwedge_{i\in PE} (~  {\tt pe}_i ~\vee~ \Not ( \bigvee_{j\in comp} pos\_j = {\tt pe}_i ) ~) 
\]

The involved controllable variables defined in the \With part will be, for each component:
{\tt 
a\_$j$},
{\tt 
b\_$j$},
$j\in comp$.
The effect of the control will be that, 
when some processing element becomes unavailable,
appropriate migrations will be fired.
More precisely:
 upon reception of input {\tt pes}$_i$ at value \True,
the availability model makes a transition as shown in Figure~\ref{fig:core-disp-autom}.
In the global model, a transition will be taken to a next state where the above expression is \True.
For components on the unavailable PE, the automata shown in Figure~\ref{fig:task-pe-autom} 
can not stay in the current state without violating the property,
therefore controllables will take a value such that a local transition occurs,
hence firing a migration action towards another available PE.

\paragraph*{Workload balancing} Workload on each PE$i$ is bounded by  $Max_i$: 
this is achieved by {\em component migration}.The expression  is:

\(
{\tt wl\_ba =} \bigwedge_{i\in PE} (~  {\tt wl}_i \leq Max_i~) 
\).

The involved controllables are:
{\tt 
a\_$j$},
{\tt 
b\_$j$},
$j\in comp$.
The effect of the control will be that, 
if some migration or starting of a component happens on a PE, 
the choice encoded by controllables 
 will be between PEs for which this addition would not violate the bound.

 \paragraph*{Quality-of-Service} When the \filehandlerone is
 overloaded, start the \filehandlertwo: this is achieved by {\em
   lifecycle and binding control}. The expressions are:

{\tt qos=\(((\Not {\tt full\_S1}) \Rightarrow ( {\tt disc\_S2} ))\)} 
\\
\(
{\tt and } ((\Not {\tt run\_L} \And {\tt full\_S1}) \Rightarrow ( {\tt run\_S2} \And \Not {\tt disc\_S2} ))
\).

The involved controllables are: {\tt s\_S2}, {\tt ch\_S2}.  The effect
of the control will be that, if the FIFO of \filehandlerone reaches
its threshold, the \filehandlertwo is started unless the Logger runs;
and when the FIFO goes back under the threshold, it is disconnected,
and eventually stopped.

\paragraph*{Exclusiveness} When the \logger runs, 
  \filehandlertwo doesn't:
this is achieved by {\em lifecycle and binding control}. The corresponding expression  is:
\[
{\tt exc =} 
({\tt run\_L} \And \Not {\tt disc\_L}) \Rightarrow {\tt disc\_S2}
\]

The involved controllable is: {\tt s\_S2}.  The effect of the control
is that, when the Logger is started, the \filehandlertwo must be stopping
or idle.
   

\section{Manager  Integration}
\label{sec-impl}

The C code generated by the BZR compilation tool-chain consists of two
functions {\em step(...)} and {\em
  reset(...)}. Listing~\ref{cod:step-call} is a sketch of a program
using these functions in order to manage the reconfiguration of a
system. At line 2, {\em reset()} is called once to initialize the
memory of the program. The piece of code ranging from line 5 to line
10 is made sensitive to incoming events; i.e., this part of the code
is executed each time new events are received. The function {\em
  prepare\_events()} prepares the events and provides the inputs for
the function {\em step()}. The signature of the function {\em step()}
corresponds to the one of the main program of
Listing~\ref{cod:main-prog}. A call to this function corresponds to a
step of the synchronous program. The outputs of {\em step()} are given
to the function {\em generate\_commands()} in order to translate the
outputs into the consequent reconfiguration commands.

\lstset{numbers=left,numberstyle=\tiny,stepnumber=1,numbersep=5pt,  basicstyle=\scriptsize}
\begin{lstlisting}[caption={Principle of manager integration}, captionpos=b,
  label={cod:step-call},language={c++}]
// program initialization
reset(&mem);
...
// reading input events
<pes1, add_L, f_S1, f_S2, f_L> = prepare_events();
// calling the step function
<outputs>=step(pes1, add_L, f_S1, f_S2, f_L, &mem);
// translate outputs to reconfiguration commands
generate_commands(<outputs>);
...
\end{lstlisting}


\subsection{Wrapping the manager into a component}
\begin{figure}[h]
\centering
\scalebox{0.8}{\input{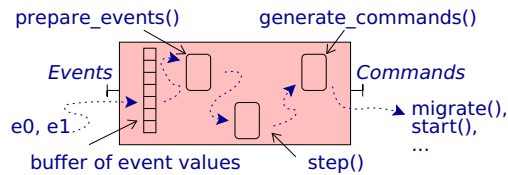}} 
\caption{The manager as a Fractal component}  
\label{fig:controller-interface}
\end{figure}

The functions listed in Listing~\ref{cod:step-call} are wrapped into a
Fractal component in order to be connected to the rest of application
and middleware components (see Figure~\ref{fig:controller-interface}).
This component provides one server interface {\em Events} and one
client interface {\em Commands}. The signature of the interface {\em
  Events} is described in Listing~\ref{cod:itf-events-comete}. The method
{\em setEvent} provided by this interface enables components to
register event values. Indeed, the manager hides a buffer containing
the last value registered of each event. The function {\em
  prepare\_events()} uses this buffer to provide the inputs for the
function {\em step()}. 

The function {\em generate\_commands()} translates the outputs of the
function {\em step()} into method calls through the {\em Commands}
interface. Indeed, the signature of the required interface {\em
  Commands} is that of {\em ComponentManager} provided by Comete API
(see Listing~\ref{cod:itf-events-comete}). It provides methods for
component migration, reconfiguration and lifecycle management.

\lstset{numbers={none}}
\begin{lstlisting}[caption={{\em Events} and {\em Commands} interface signature}, captionpos=b, label={cod:itf-events-comete},language={idl}]
interface SynchronousManager.api.Events{
  setEvent(in event_name, in event_value)
}

interface Comete.generic.api.ComponentManager {
  instantiate(in component_binary_name, 
              in target_pe_id, out instance_id)
  destroy(in instance_id)
  bind(in client_id, in client_interface_name, 
       in server_id, in server_interface_id, 
       in binding_type)
  unbind(in client_id, in client_interface_name)
  start(in instance_id)
  stop(in instance_id)
  migrate(in instance_id, in target_pe_id)
}


\end{lstlisting}






\subsection{Concrete integration of the manager}

\begin{figure}[h]
\centering
\input{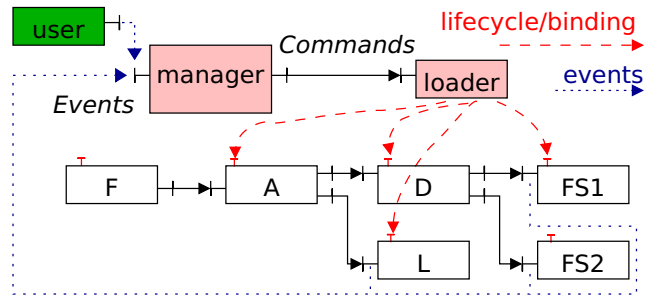}
\caption{Integration of the manager in the system} 
\label{fig:controller-integration}
\end{figure}

Figure~\ref{fig:controller-integration} describes the concrete
integration of the manager together with the application and Comete
middleware components. The component {\sf user} is connected to the
{\em Events} interface of the manager. It intercepts user requests
(add logger for instance) and registers them as events. The FIFO
components (part of Comete) are made implicit in the figure. As
illustrated by the dotted lines, the FIFOs associated with
\filehandlerone, \filehandlertwo and \logger are connected to the {\em
  Events} interface of the manager in order to register events related
to their size.

Upon receiving events, the manager performs some computation and
fires some reconfiguration commands in the form of method
calls. These method calls are handled by the {\em loader} of
Comete. Indeed, as illustrated by the dashed lines, the {\em loader}
is connected to the control interface of the application components in
order to start/stop bind/unbind them. Moreover, the loader knows how
migrating components should be achieved regarding the execution
platform.

Comete middleware hides complex mechanisms and provides a simple API
to manage component lifecycle, migration, and architecture
transparently, whatever the execution platform is. For that purpose,
we adopted a centralized version of the \syncprog emitting 
reconfiguration commands to Comete instead of application
components. We could make the \syncprog communicate directly with the
components (modulo re-engineering Comete) to enable a distributed
implementation of the \syncprog. This means that parts of the
\syncprog would be integrated into Fractal component membranes.

In this section we presented a simple version of the use of a
synchronous manager for the reconfiguration of a parallel and
asynchronous application. The reconfiguration commands were considered
to be short enough to complete before the next reaction of the
manager. However, this is not always the case. Indeed, some
reconfiguration commands may take non negligible time to
complete. During this time interval, the system model inside the
manager does not reflect the actual state of the system.  For
instance, when the command to start a component {\em C$_x$} is fired,
the model of {\em C$_x$} is at state {\tt Running (R)} but the actual
component is not yet started. The manager reaction to other incoming
events during reconfiguration progress may lead the managed system to
undesirable states. Next section proposes an extension to our
modeling approach in order to solve issues related to asynchronous commands.





\section{Asynchronous 
commands}
\label{sec-impl-async}



We propose a controller architecture together with some guidelines for
writing synchronous models in order to overcome the issues related to
the transient incoherence of the model with respect to the state of
the system. Our approach relies on the explicit representation of the
states where some commands are being processed but not completed,
together with some synchronization mechanisms.

The purpose of identifying the states where some reconfiguration
commands are in progress is to make it possible for the programmer to
decide what should happen during reconfigurations. The programmer
should rewrite objectives properties taking into account these
situations.

\subsection{Modeling command execution}

\subsubsection{Behavioral models}

Some synchronous languages (e.g., Esterel) provide built-in constructs
in order to perform asynchronous calls within a synchronous
program~\cite{berry-popl93,paris-thesis}. We follow the same principle
of such constructs and model the asynchronous\footnote{by asynchronous
  we mean that a command takes more than one reaction step to
  complete.} aspect of reconfiguration commands by means of an
automaton.

\begin{figure}[h]
\centering
\input{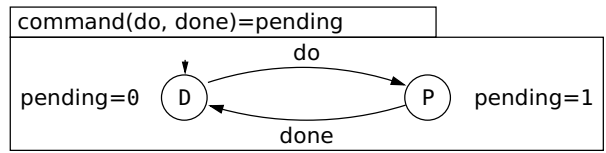}

\caption{Automaton for an asynchronous action.}
\label{fig:command-autom}
\end{figure}

Such an automaton is illustrated by Figure~\ref{fig:command-autom}.
At state \Spend (for pending), the command is emitted but not yet
completed. At state \Sdone, the command has finished. The automaton
has two inputs {\sf do} and {\sf done}. The output {\sf pending} tells
whether the command is pending or not. From the initial state, upon
receiving {\sf do} which corresponds to the firing of the command, the
automaton goes to state \Spend. The automaton stay at state \Spend
until receiving the input {\sf done}.  The input {\sf done} signals
that the command has been completed.

We associate an instance of such an automaton with each asynchcronous
command.  Hence the modification w.r.t. model of
Section~\ref{sec-ctrl} is systematic and modular.

\subsubsection{Objectives and contracts}

Being aware of the state where a command is in progress but not
finished, one can add some properties to be enforced and 
modify the previous ones in order to decide what should happen during
reconfigurations. For illustration purpose, we consider the same
program as in Section~\ref{ss-controller-main} with the command {\sf
  start\_S2} as the only command taking non-negligible time to
complete. In the following fragment of program, we associate an
instance of the node {\tt command} to {\sf start\_S2} (line 5) and modify the
contract:

\lstset{numbers=left,numberstyle=\tiny,stepnumber=1,numbersep=5pt}
\begin{lstlisting}[language={[bzr]c++}, mathescape]
...
enforce (pend and  pe_av and wl_ba and 
         (not pen_sS2 $\implies$ (qos and exc)))
...
pen_sS2 = command(start_S2, done);
pend = not (pen_sS2 and disc_S2) ; ...
\end{lstlisting}

The contract changes as follows:
\begin{compactitem}
\item The properties {\sf qos} and {\sf exc} are now dependent to the
  pending of the command. \texttt{\Not pen\_sS2$\implies$(qos \And exc)}
  tells that {\sf qos} and {\sf exc} may not be enforced when the
  command {\sf start\_S2} is in progress. But, they must be enforced
  once the command completes.
\item {\sf pend} (defined at line 6) is a new property. It forbids the
  controller to modify the state of the \filehandlertwo when the
  command to start it is in progress. This property is explained
  below.

\end{compactitem}

\begin{figure}[h]
\centering
\input{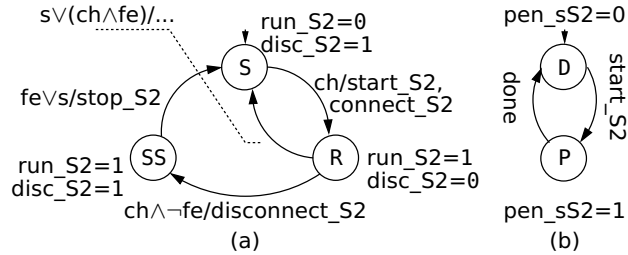}

\caption{Instances of lifecycle(a) and command(b)}
\label{fig:product-life-pend-autom}
\end{figure}

In order to understand the property \textsf{pend}, consider the automata
modeling \filehandlertwo lifecycle and {\tt start\_S2} command
execution in Figure~\ref{fig:product-life-pend-autom}. Adding the
property \mbox{\texttt{pend = \Not (pen\_sS2 \And disc\_S2)}} forces the
controller to avoid the states where the command is pending (i.e., at
state {\sc p}) and any state where {\sf disc\_S2} is true (i.e., at
states {\sc s} and {\sc ss}). That is, to avoid the states (\Sssp) and
(\Ssp) in the product of the two automata.

At the beginning the automata are at state (\Ssd). Upon receiving {\sf
  ch}, automaton (a) changes states and fires the command {\sf
  start\_S2} which makes automaton (b) change state at the same
instant. That is, the global state is (\Srp). Unless receiving {\sf
  done}, which is uncontrollable, the controller will not make
automaton (a) change state in order not to violate the property {\sf
  pend}.

\subsection{Controller Architecture}

\begin{figure}[h]
\centering
\input{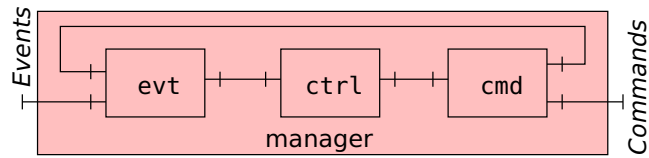}

\caption{Internal view of the controller}
\label{fig:controller-arch}
\end{figure}

The actual implementation of the manager is
component-based. Figure~\ref{fig:controller-arch} describes the
internals of such a component. It consists of three subcomponents:
\evtreceiver, \synctrl, and \cmdmanager.  The functions {\em
  prepare\_events()}, {\em reset()}, {\em step()}, and {\em
  generate\_commands()} are spread over the subcomponents as follows:
\evtreceiver encapsulates {\em prepare\_events()}. It prepares the
inputs to \synctrl that encapsulates {\em reset()} and {\em step()}
generated by the BZR compilation. The component \cmdmanager translates
the outputs of \synctrl into reconfiguration methods calls.

The component \evtreceiver is in charge of implementing the method
provided by the interface {\em Events} from which components and the
environment may register events. The component \cmdmanager is in
charge of calling reconfiguration methods through {\em Commands}
interface. Moreover, each time a reconfiguration command is completed,
\cmdmanager notices {\tt done} events related to this command. That is
why \cmdmanager is bound to \evtreceiver.

\section{Related work}

  A lot of work has been devoted to dynamic reconfiguration of
  component-based software systems. In the case of Fractal, one can
  refer to the tools for component introspection and the languages for
  specifying reconfigurations~\cite{fscript-fpath}
or integration in
parallel frameworks
\cite{BuiAndPaz05PARCO}.

Our work fits in the context of applying formal methods for dynamic
reconfiguration~\cite{beh-fractal-dis,reliable-reconf, LTL-reconf,
  bip-fractal} in order to ensure properties related to components and
reconfigurations. In~\cite{LTL-reconf} the authors specify
reconfiguration properties by means of temporal logic in order to
apply model checking or runtime monitoring of system
reconfigurations. In comparison, our approach benefits from the
discrete controller synthesis which provides a correct by construction
manager ensuring properties on components and reconfigurations.
%
In~\cite{cbse10}, the authors give a general software engineering framework, with
some indications of integration, but no complete implementation.  In
comparison, we perform a concrete integration of synchronous managers
with Fractal, in the form of its Cecilia programming environment, and
the Comete middleware.  In addition, we consider and treat the case of
asynchronous reconfiguration actions.

Other related work concern the concrete integration of synchronous
programs for the management of asynchronous systems. In \cite{BMM11},
the authors provides a synchronous controller for configuring device
drivers aiming at global power management of embedded systems. Apart from
their not  using DCS techniques, the difference with our approach lies in
the call of reconfiguration functions. Indeed, in~\cite{BMM11},
function calls are performed inside the reaction of the
controller. This has the benefit of always keeping a model reflecting
the exact state of the system. In our case we use to follow the
principle of asynchronous function calls as in~\cite{berry-popl93,
  paris-thesis}.

\section{Conclusion}



{ Our contribution is on the one hand a synchronous model of the
  behavior of the reconfigurable components, in the form of the state
  space of the control problem, and the specification of the control
  objectives (Section~\ref{sec-ctrl}).  On the other hand, we
  contribute a component-based architecture for the implementation
  of the resulting controller with the Fractal/Cecilia programming
  environment, taking advantage of the Comete
  middleware. (Sections~\ref{sec-impl} and ~\ref{sec-impl-async}) }

In this work, we apply formal techniques issued from academic research
to an open, stable and available framework supported by an independent
industry consortium \cite{ow2-website}.  The Fractal model is
currently being implemented in the \MIND project \cite{ow2-mind}, a
sequel to the Cecilia framework.  The \MIND framework is a
collaborative initiative to spread component-based software
engineering to an even larger community of academic institutions and
industries. The present contribution keeps its legitimacy in the \MIND
context as most of the concepts and semantics introduced with Cecilia
and Comete remain.  Therefore, applying our contribution to
\MIND should be limited to a straightforward adaptation
.


{ We have ongoing work on another interesting case study: an H264
  video processing application, implemented on a multicore
  architecture using Cecilia and Comete. It could be handled following
  the very same methodology we propose, because it follows essentially
  the same structure: degrees of reconfiguration concern migrations,
  and adding and removal of components performing video effects on the
  stream.  }

Perspectives are in the line of 
generalizing our proposal at the language-level, by extending the Fractal  ADL with 
a way to incorporate automata notation for reconfiguration description, as well as the behavioral contracts,
and integrating the application of BZR in a global Fractal compilation flow.
Also, the integration of synchronous controllers in a Fractal components architecture
should be facilitated by identifying general programming guidelines,
providing end-users with informal rules such that
components are controllable,
and
the synchronous instant and step (and the states in the automata) fits well with the event granularity.


\newpage
\appendix 

\section{Complete Example and Integration}  

In what follows, we describe the complete integration process of the
code produced by the {\sc bzr} compilation for the control of the {\em
  comanche} application software. Writing the synchronous model of the
manager has been discussed above. Here we give more details providing
simulation results of the manager and the process of integration in
the Fractal application.

\section{Simulation of the Synchronous Program}
\label{s-simu}
The interface of the synchronous program is described in
Listing~\ref{cod:node-main}. The input {\em disable} states on the
availability of the processing element {\em PE$_3$}. This input takes
the value {\em false} as long as {\em PE$_3$} is available ({\em false 
  otherwise}). {\em fifoH1F, fifoH2F, fifoL2F} are Booleans stating on
the size of the FIFOs associated with FileServer1, FileServer2 and
Logger respectively. They take the value {\em false} as long as the
size of the associated FIFOs is under their specific threshold. The
input {\em addlog} tells whether the logger is required or not. The
input {\em c\_startH2\_done} notifies that the command for starting
FileServer2 has been actually completed.

The outputs of the main program are associated with the possible
commands provided by the reconfigurable system. An output associated
with a command takes the value {\em false} until the command has to be
fired. Then, it takes the value {\em true} during one step.

\begin{lstlisting}[caption={The interface of the main program}, 
captionpos=b, label={cod:node-main},language={[bzr]c++}]
node main(disable, fifoH1F, fifoH2F, fifoL2F, addlog, c_startH2_done:bool) 
  returns(
    c_startH2, c_stopH2, c_connectH2, c_diconnectH2:bool;
    c_startL, c_stopL, c_connectL, c_diconnectL:bool;
    c_mig1f1 , c_mig2f1 , c_mig3f1 , c_mig4f1,
    c_mig1f2 , c_mig2f2 , c_mig3f2 , c_mig4f2 ,
    c_mig1d , c_mig2d , c_mig3d , c_mig4d:bool;
    ...)
\end{lstlisting}

Figure~\ref{fig:simu} illustrates part of the simulation results. The
inputs (in blue color) are given (by the user) in order to simulate
the events received from the application. The outputs (in red color)
corresponding to reconfiguration commands take the value {\em true}
each time a command has to be fired.

\begin{figure}[h]
\centering
\input{simu}
\caption{}  
\label{fig:simu}
\end{figure}

The interaction scenario as illustrated by the figure is as follows:
\begin{itemize}
\item Step 1 is for initializing the system (not relevant).
\item at Step 4  the input fifoH1F changes value to {\em true}. The program
  reacts and assigns the value {\em true} to the outputs {\em c\_startH2} and
  {\em c\_connectH2} to true. These are the commands to be fired at
  this step. This situation corresponds to the activation of
  \filehandlertwo when \filehandlerone is overloaded.

\item at Step 8, the program receives the notification that {\em
    c\_startH2} has been completed.

\item at Step 10 the input fifoH2F changes value to {\em true}. The
  program reacts but no command is fired. 

\item at Step 13 the input fifoH1F changes value to {\em false}. The
  program reacts and assigns the value {\em true} to the output {\em
    c\_disconnectH2}. This corresponds to the situation where the
  FileServer2 is no longer required. It should be disconnected and
  left executing in order to terminate the remaining inputs in its
  FIFO.

\item at Step 16, the input fifoH2F changes value to {\em false}. The
  program reacts and assigns the value {\em true} to  {\em
    c\_stopH2}. Now that the FIFO associated with the FileServer2 is
  empty, the component can be completely stopped.

\item at Step 18, the input disable changes to {\em true}. The program
  reacts and assigns the value true to {\em c\_mig1f2} and {\em
    c\_mig1d}. This corresponds to the migration of the components
  running on {\em PE$_0$} to other {\em PE$_s$}. The output {\em
    c\_mig1f2} (resp., {\em c\_mig1d}) corresponds to the migration of
  the \filehandlertwo (resp., \dispatcher) to {\em PE$_1$}.
\end{itemize}

\subsection{BZR Program Compilation into C Code}
The BZR compilation tool chain compiles the synchronous program into C
code. The generated functions of our interest are the reset() and
step() functions of the main program. reset() initializes the memory
associated with the synchronous program. The function step() computes
the outputs of the program depending on the inputs. It also updates
the internal memory.

\begin{lstlisting}[caption={Prototype of the function step}, 
captionpos=b, label={cod:proto-step},language={c++}]

main_res main_step(int disable, int fifoH1F, int fifoH2F,
                   int fifoL2F, int addlog,
                   int c_startH2_done, main_mem* self);
                                         
\end{lstlisting}

The prototype of the function step() is presented in
Listing~\ref{cod:node-main}. It has the same input parameters as the
main function of the synchronous program in
Listing~\ref{cod:node-main}. The parameter {\tt self} refer to the
memory of the complete program. The returned value of the function step() is
of type {\tt main\_res} it is a structure. Each field of this structure
corresponds to an output of the main program.

\section{Integration of the step() function}

The integration process of a synchronous manager for the control of
reconfiguration lies in the integration of the step() function
generated by the synchronous compilation. In order to achieve this, we
have to make some choices. In particular, we need to decide when the
function should be called, and how to apply reconfiguration
commands:
\begin{itemize}
\item \textbf{Calling Step():} Two approaches are possible:
  Time-triggered or Event-triggered. The first solution consists of
  calling the step periodically each {\tt t time units}. The second
  one consists of making the step function sensitive to incoming
  events; i.e., each time a new event is notified, step is called.
\item \textbf{Firing commands:} The outputs of the step function
  correspond to commands to be fired. One way to apply these commands
  is to call the reconfiguration commands before the step is
  finished. The advantage of doing so is to have a model always
  reflecting the state of the system. The other approach is to
  complete the step and deal with the commands to be fired in an
  asynchronous way.
\end{itemize}

Our choice for this case study is to consider event-triggered calls of
the step together with asynchronous firing of commands. The choice may
be discussed. But this is out of the scope of the paper.

\subsection{Wrapping The Generated Code Into a Fractal Component}
The C generated code is wrapped into a Fractal component in order to
receive application events and fire reconfiguration
commands. Figure~\ref{fig:manager} describes the internals of such a
component and the type of the provided (resp. required) interface of
the components. Notice that two interfaces may be bound if an only if
they are of the same type.

\begin{lrbox}{\loader}
\begin{lstlisting}[framexrightmargin=-6.5cm, language={idl2}]
interface manager.api.commands {
  void migrate(in cpt, in core);
  void start(in cpt);
  void stop(in cpt);
  void bind(in cpt);
  void unbind(in cpt);
}
\end{lstlisting}
\end{lrbox}

\begin{lrbox}{\cmd}
\begin{lstlisting}[language={idl2}, framexrightmargin=-7cm]
interface Actions{
  void doAction(in commands);
}
\end{lstlisting}
\end{lrbox}

\begin{lrbox}{\evt}
\begin{lstlisting}[language={idl2}, framexrightmargin=-6cm]
interface manager.api.events{
  void setValue(in event, in value);
}
\end{lstlisting}
\end{lrbox}

\begin{lrbox}{\sync}
\begin{lstlisting}[language={idl2},framexrightmargin=-8cm]
interface Synchronous{
  void step(in inputs);
}
\end{lstlisting}
\end{lrbox}

\newcommand{\eventsapi}{%
\begin{minipage}{8cm}
\usebox\evt 
\end{minipage}%
}%

\newcommand{\loaderapi}{%
\begin{minipage}{8cm}
\usebox\loader
\end{minipage}%
}%
\newcommand{\cmdapi}{%
\begin{minipage}{8cm}
\usebox\cmd
\end{minipage}%
}%
\newcommand{\syncapi}{%
\begin{minipage}{8cm}
\usebox\sync
\end{minipage}%
}%

\begin{figure}[h]
\centering
\input{manager}
\caption{}  
\label{fig:manager}
\end{figure}

\subsubsection{The  Component EventReceiver}

Listing~\ref{cod:event-rec} illustrates part of the implementation of
the method {\em setValue} of the {\tt EventReceiver} component. This
method is called by other components in order to notify events.

The component manages an array of Boolean values
(DATA.step\_inputs). Each cell of the array corresponds to one input  
event. Each time an event is notified through this method, its value
is updated, then the method step is called. This method is implemented
by the component {\tt SynchronousProgram}. The parameter of the call
is the array of event values. 

\begin{lstlisting}[caption={code of the setValue method}, 
captionpos=b, label={cod:event-rec},language={c++}]
void METHOD(events, setValue)(void *_this, int event , int value) {
  ...
  DATA.step_inputs[event]=value; 
  ...
  CALL(REQUIRED.synchprog, step, DATA.step_inputs);
}

\end{lstlisting}

\subsubsection{The  Component SynchronousProgram}

Listing~\ref{cod:sync} illustrates the code associated with the method
{\em step} of the component {\tt SynchronousProgram}. This method
contains a call to the {\em step} method actually generated by the BZR
compilation (at line 7). The result of the synchronous step (i.e., {\tt DATA.res})
are forwarded to the {\tt CommandsGenerator} in order to translate them
into the consequent reconfiguration commands.

\begin{lstlisting}[caption={Code of the step method}, 
captionpos=b, label={cod:sync},language={java}]
void METHOD(step, step)(void *_this, jboolean *inputs) { 
  int input_disable = inputs[0];
  int input_fifoH1F = inputs[1];
  ...
 
  //Call to the generated method step()
  DATA.res = Migration_main_step(input_disable ,input_fifoH1F ,
				 input_fifoH2F , input_fifoL2F, 
				 input_addlog ,input_c_startH2_done, 
				 &DATA.mem);
  
  CALL(REQUIRED.actionManager, doAction, (void *) &DATA.res);
  }
\end{lstlisting}

\subsubsection{The  Component CommandsGenerator}

Listing~\ref{cod:cmd-gen} illustrates the implementation of the method
{\em doAction} of the component {\tt CommandsGenerator}. It receives a
structure containing the Boolean values associated with the possible
commands. This method parses the received values and fire a command
each time the associated output variable has the value {\tt true}.

\begin{lstlisting}[caption={Code of the doAction method}, captionpos=b, label={cod:cmd-gen},language={java}]
void METHOD(actionManager, doAction)(void *_this, void * sync_output) {  
  Migration_main_res *commands;
  commands = (Migration_main_res *)sync_output;

  if(commands->Migration_c_startH2)
    CALL(REQUIRED.commands, start, FILES2);  

  if(commands->Migration_c_stopH2)
    CALL(REQUIRED.commands, stop, FILES2);
 ....
\end{lstlisting}
\section{Retrieving Application/Environment Events}

As explained above, the component {\tt EventReceiver} of the manager
offers a method for event notification. The components in charge of
notifying these events are connected to this interface. Among the
application components, only the FIFOs associated with
\filehandlerone, \filehandlertwo and \logger are supposed to send
events to the manager. Moreover, there is a component modeling the
user and the environment. The component is in charge of notifying the
events related to adding the logger and the availability of a
processing element. These components call the method {\em setValue()}
of the component {\tt EventReceiver} whenever a particular condition
holds. Listing~\ref{cod:evt-not} is a sample piece of program for
notifying events. The condition depends on the nature of the event to
be notified. In case of the FIFOs, the condition concerns the size of
the FIFO regarding to its associated threshold.
    
\begin{lstlisting}[caption={Event notification}, captionpos=b,
  label={cod:evt-not},language={java}]
  ...
  if(condition)
  CALL(REQUIRED.events, setValue, EVENT_ID, EVENT_VALUE);
  ...
\end{lstlisting}
  
\bibliographystyle{abbrv}    
\bibliography{emsoft}
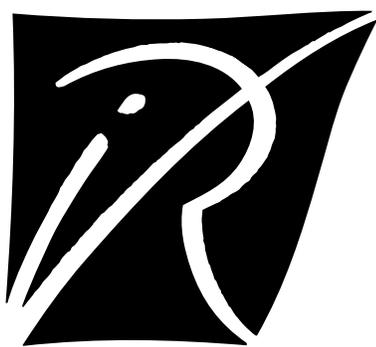
\end{document}